\documentclass[a4paper]{cas-sc}

\usepackage[english]{babel}
\usepackage{hyperref}
\usepackage{amsthm}
\usepackage{soul}
\usepackage{amsmath}
\usepackage{mathrsfs}
\usepackage{euscript}

\usepackage{algorithm}
\usepackage[noend]{algpseudocode}
\usepackage{ragged2e}
\usepackage{makecell}
\usepackage[numbers]{natbib}
\usepackage{booktabs}
\usepackage{graphicx}
\usepackage{subcaption}
\usepackage{pifont} 
\usepackage{lineno}
\usepackage{xcolor, soul}

\def\tsc#1{\csdef{#1}{\textsc{\lowercase{#1}}\xspace}}
\tsc{WGM}
\tsc{QE}
\tsc{EP}
\tsc{PMS}
\tsc{BEC}
\tsc{DE}

\begin{document}
\let\WriteBookmarks\relax
\def\floatpagepagefraction{1}
\def\textpagefraction{.001}

\shorttitle{Empirical Validation of Conformal Prediction}

\shortauthors{J. Fayyad et~al.}

\title [mode = title]{Empirical Validation of Conformal Prediction for Trustworthy Skin Lesions Classification}                      



%
\author[1,3]{Jamil Fayyad} [orcid=0000-0003-1553-8754]


\ead{jfayyad@mail.ubc.ca}



\affiliation[1]{organization={The University of British Columbia},
    addressline={3333 University Way}, 
    city={Kelowna},
    postcode={V1V 1V7}, 
    state={BC},
    country={Canada}}

\author[2,3]{Shadi Alijani}[]


\ead{shadialijani@uvic.ca}


\affiliation[2]{organization={University of Victoria},
    addressline={800 Finnerty Road}, 
    city={Victoria},
    postcode={V8P 5C2}, 
    state={BC},
    country={Canada}}

\author[2,3]{Homayoun Najjaran}
\cormark[1]
\ead{najjaran@uvic.ca}

\affiliation[3]{organization={Cognia AI},
    addressline={2031 Store street}, 
    city={Victoria},
    postcode={V8T 5L9}, 
    state={BC},
    country={Canada}}


\cortext[cor1]{Corresponding author}



\begin{abstract}
\\
\noindent \textit{Background and objective:} Uncertainty quantification is a pivotal field that contributes to realizing reliable and robust systems. It becomes instrumental in fortifying safe decisions by providing complementary information, particularly within high-risk applications. existing studies have explored various methods that often operate under specific assumptions or necessitate substantial modifications to the network architecture to effectively account for uncertainties. The objective of this paper is to study Conformal Prediction, an emerging distribution-free uncertainty quantification technique, and provide a comprehensive understanding of the advantages and limitations inherent in various methods within the medical imaging field.

\noindent\textit{Methods:} In this study, we developed Conformal Prediction, Monte Carlo Dropout, and Evidential Deep Learning approaches to assess uncertainty quantification in deep neural networks. The effectiveness of these methods is evaluated using three public medical imaging datasets focused on detecting pigmented skin lesions and blood cell types. 

\noindent\textit{Results:} The experimental results demonstrate a significant enhancement in uncertainty quantification with the utilization of the Conformal Prediction method, surpassing the performance of the other two methods. Furthermore, the results present insights into the effectiveness of each uncertainty method in handling Out-of-Distribution samples from domain-shifted datasets. Our code is available at:

\noindent\textit{Conclusions:} Our conclusion highlights a robust and consistent performance of conformal prediction across diverse testing conditions. This positions it as the preferred choice for decision-making in safety-critical applications.

\end{abstract}




\maketitle

\section{Introduction}

Deep learning algorithms have demonstrated significant potential in solving a wide range of complex, real-world problems \cite{fayyad2020deep}. These algorithms' capacity to autonomously learn intricate patterns and representations makes them valuable in diverse computer vision tasks, including image classification \cite{wiley2018computer, alijani2022ensemble}, object detection \cite{salari2022object}, and remote sensing \cite{zhu2017deep}. Generally, the evaluation of deep learning algorithms relies on accuracy as the main metric to assess how well the model has learned from the training data. While accuracy is widely acknowledged in the literature, it alone does not guarantee the system's robustness. This is more apparent in safety-critical applications where errors in decision-making can have catastrophic consequences \cite{lambert2022trustworthy}. To mitigate the risk of erroneous decisions, presenting the confidence level associated with predictions made by deep learning algorithms is imperative. For this purpose, uncertainty quantification techniques are instrumental in providing models with additional information that presents prediction confidence and ultimately helps improve the model's robustness.

In medical applications, deep learning plays a vital role by enabling more accurate and efficient analysis of complex data, including medical image classification, tumor detection, and medical image segmentation. Deep learning models, however, are trained on clean and well-curated datasets, which enables models to achieve high accuracies on such data. However, images are often subjected to noise, artifacts, and inherent limitations of the imaging process. These factors can potentially impact the predictions of deep learning models, leading to uncertainty in their outputs. Therefore, it is crucial to express confidence in the network's predictions to assess the reliability and robustness of its results, especially in critical applications such as medical imaging.

In machine learning, two sources of uncertainties collectively influence model predictions \cite{fayyad2023out}. The first source, known as epistemic uncertainty, arises from a lack of knowledge and understanding. Model retraining and incorporating additional information and data are typically employed to mitigate this type of uncertainty \cite{ghesu2021quantifying}. The second source of uncertainty is aleatoric uncertainty, which characterizes the inherent randomness within the data and, as such, remains irreducible. In the existing literature, various approaches offer both theoretical and practical solutions for quantifying uncertainties associated with deep learning models \cite{yan2023uncertainty, jahmunah2023uncertainty}. These methods can be broadly categorized into \textit{Single Deterministic Networks}, \textit{Bayesian Neural Networks}, \textit{Deep Ensembles}, and \textit{Test-time Augmentation methods} \cite{abdar2021review}. 

Single deterministic networks are straightforward methods to represent the predictive uncertainty of deep learning models. As suggested by the name, the networks are trained to adjust their weights in a deterministic manner, and the uncertainty is measured through a single forward pass. A simple baseline of single deterministic networks is the utilization of the neural network output probabilities (i.e., SoftMax outputs for classification) as an interpretation of the model uncertainty. Several studies, however, showed that neural networks are often over-confident \cite{wei2022mitigating}, hence their output probabilities are miscalibrated \cite{guo2017calibration} and tend to generate wrong uncertainty estimates. A more generic approach in single deterministic networks includes predicting the parameters of a prior distribution over the outputs. such algorithms are known as Evidential learning \cite{sensoy2018evidential} or Prior Networks \cite{malinin2018predictive}. The framework of these approaches involves introducing a higher-order distribution over the likelihood function, with the parameters of these distributions being estimated using the underlying neural network. The predicted class probability corresponds to the expected value of the distribution, while the variance of the distribution serves as an estimate of the model's uncertainty. In \cite{feng2024trusted}, Feng et al. proposed a vision transformer-based backbone as a reliable multi-scale classification framework for Whole Slide Images. The backbone is integrated with evidence uncertainty theory to assess uncertainty levels at different magnifications of a microscope. Despite the demonstrated outperforming results, evidence-based approaches require changes to the model architecture and the loss function employed at training. 

Bayesian Neural Networks (BNNs) \cite{kononenko1989bayesian} seamlessly merge the principles of Bayesian inference with traditional neural networks. Instead of considering BNN parameters as fixed, deterministic values, they are treated as probability distributions. This unique approach empowers BNNs to effectively capture and represent predictive uncertainty. Consider a neural network with parameters $\omega$ and a training dataset $D_t=(x, y)$. If the network weights are assigned a prior distribution, denoted as $P(\omega)$, the posterior distribution $P(\omega | x, y)$ can be calculated using Bayes' theorem. To calculate the posterior $p(\omega | X, Y)$, it necessitates integrating over all possible values of $\omega$, a task that is often intractable. Generally, obtaining the posterior is only viable through either Variational Inference \cite{blei2017variational}, where a simple and tractable distribution is used to approximate the exact posterior, or through other approximation techniques, such as Markov Chain Monte Carlo (MCMC) \cite{brooks1998markov}. Leibig et al. \cite{leibig2017leveraging} utilized MCDropout to assess the uncertainty of deep learning models in detecting diabetic retinopathy. while the method captures useful uncertainty information, the sampling process is computationally expensive, hence it hinders real-time implementation.

Deep ensembles involve training several deep networks with random initializations on a given dataset and deriving a unified prediction from the collective outputs of the ensemble members \cite{lakshminarayanan2017simple}. Although initially designed to improve prediction accuracy, deep ensembles have shown their capacity to effectively estimate predictive uncertainties in various studies. Consider a \(M\) number of independent networks that map the inputs \(x\) to a label \(y\), such that \(y = f_{\theta_i}(x)\), each network is parameterized by \(\theta\) parameters, and \(i \in 1, \ldots, M\). The combined prediction of the ensemble can be simply represented as the average predictions of each network at inference.

Recently, Conformal prediction (CP) \cite{shafer2008tutorial, vovk2005algorithmic} emerged as a prominent uncertainty quantification technique. It has gained attention in the fields of computer vision \cite{angelopoulos2020uncertainty,angelopoulos2022image}, and more specifically, in medical imaging \cite{lu2022fair}. In addition to its elegance and computational efficiency, CP algorithms belong to the domain of 'distribution-free' methodologies. This implies that they are model-agnostic, data distribution agnostic, and applicable to finite sample sizes. 

In this paper, we compare three top-performing uncertainty quantification methods, namely CP, Monte Carlo Dropout (MCD), and Evidential Deep Learning (EDL), and provide a comprehensive and detailed analysis of their performance. The selection of these three methods is attributed to the fact that they cover diverse approaches, including Bayesian, evidential, and distribution-free methodologies. Furthermore, our selection process of the algorithms encompassed a range of implementation details, spanning from maintaining the original network structure without modification to undertaking complete retraining and adjusting the loss functions. The central research questions under examination are: \textit{How stable and robust are the prominent uncertainty quantification methods, and how do they perform individually in real environment settings?}
We include three experiments to underscore the unique attributes of each method, with a particular emphasis on CP, a less-explored approach for skin lesion classification tasks. The main contributions of our work are as follows:

\begin{enumerate}
    \item We proposed in-depth explorations and analysis of three deep-learning uncertainty quantification techniques, including the prominent emerging CP, for the classification task of skin lesions.

    \item We presented a comparative analysis and comprehensive evaluation of the implication of CP parameter variations. Particularly, we studied the effect of changing the scoring function, the confidence level, and the calibration set size through a detailed set of requirements.

    \item We conducted an assessment of the effectiveness of each method in addressing Out-of-Distribution (OOD) data arising from covariate shifts.  
    
\end{enumerate}

The remainder of this paper is organized as follows: Section \ref{sec:2} presents background on uncertainty quantification techniques and the related work to our study. Section \ref{sec:3} presents the experiment settings, the experiment results, and the discussions. Section \ref{sec: sec4} concludes the paper and proposes future work directions.

\section{Methods}
\label{sec:2}

\subsection{Dataset} \label{ssec:3.1}
This paper focuses on the challenge of classifying different types of skin lesions by leveraging deep-learning approaches and medical images. For this purpose, we work on three publicly available medical datasets, two skin lesion datasets named: the HAM10000 dataset \cite{tschandl2018ham10000}, and the Dermofit (DMF) and Blood Cell Microscope (BCM) \cite{acevedo2020dataset} dataset.  \cite{ballerini2013color}. The HAM10000 dataset contains more than 10,000 images and the DMF contains more than 1200 images, both captured from real patients with seven distinct lesion classes. Both datasets are highly imbalanced, with some classes having more samples than others. To address biases stemming from imbalanced classes, we initially identified the class with the highest sample count. Utilizing this count as a benchmark, we calculated the additional samples required for each class to achieve balance. Subsequently, various data augmentation methods, including vertical and horizontal flipping, angle rotations, and cropping, were employed to generate the necessary number of samples for each class.
The BCM dataset contains around 17,000 peripheral blood cell images labeled with 8 different blood cell samples.
To train the network, we divided the dataset into 50\% training, 20\% validation, and 30\% for testing. For CP, we reserved a portion of samples from both validation and testing datasets that were not used before and utilized them for calibration.

\subsection{Model Specifications and Training configuration}
\label{Sec: model specs}
In all our main experiments, we employ ResNet-18 \cite{he2016deep} as the underlying core network for all uncertainty quantification methods. The network, which is pretrained on ImageNet, consists of 18 layers that include convolutional layers, batch normalization, dropout layers, ReLU activation functions, and residual blocks. In addition, we have also utilized two additional networks: ResNet-50 and VGG-11 to study the effect of changing the core network. In all cases, a classification head is modified to classify the input images into their corresponding classes. We trained the core model using the cross-entropy loss for 100 epochs. The Adam optimizer was used with a learning rate of 1e-4. A learning rate scheduler was added to reduce the learning rate by 0.5 when the validation loss is not improved for 10 consecutive epochs. All the experiments are run on an NVIDIA RTX A5000. 

\subsection{Uncertainty Quantification Methods} 
In the rich domain of uncertainty quantification, understanding the nuances of existing methodologies and their implementation details is paramount. This section delves into the background literature and related work that forms the foundation for uncertainty methods. We specifically focus on three quantification techniques that form the basis of our experimental analysis. For a more comprehensive review of uncertainty quantification methods in deep learning, interested readers can refer to \cite{abdar2021review, caldeira2020deeply}. 
\subsubsection{Monte Carlo Dropout}
Dropout is a common practice during training to prevent overfitting, involving the random deactivation of neurons in the network. In their research, Gal et al \cite{gal2016dropout}. demonstrated that dropout can also serve as a Bayesian approximation, enabling the prediction of the network's uncertainty when applied during inference. A straightforward method to estimate the network's uncertainty involves enabling dropout during inference and conducting multiple forward passes of the same input. The random deactivation of neurons during this process introduces variability in the outputs, resulting in both a mean and a variance. The mean value serves as the prediction, while the variance serves as an estimate of the prediction's uncertainty. 

Let \( f(x) \) be the prediction of the neural network for input \( x \). During MCD, we obtain \( T \) samples of predictions \( f_t(x) \) where \( t = 1, 2, ..., T \) represents each Monte Carlo sample. These predictions are obtained by performing $T$ forward passes through the network. The mean prediction can be represented as:

\begin{align}
    \hat{f}(x) &= \frac{1}{T} \sum_{t=1}^{T} f_t(x)
\end{align}
The uncertainty can be expressed as the variance of these predictions:

\begin{equation}
    \text{\textit{Var}}(x) = \frac{1}{T} \sum_{t=1}^{T} \left( f_t(x) - \frac{1}{T} \sum_{t'=1}^{T} f_{t'}(x) \right)^2
\end{equation}

\subsubsection{Evidential Deep Learning}

In a traditional Convolution Neural Network (CNN), label probabilities are generated by applying a SoftMax activation function to the last layer. The objective is to maximize the likelihood function by minimizing a metric of loss, often a cross-entropy loss, which tends to make overconfident predictions \cite{grabinski2022robust}. The output of the SoftMax operation reflects discrete point estimates of class probabilities without proper indication of the uncertainty present in the estimate. EDL on the other hand accounts for the uncertainties in prediction by estimating output distributions instead of rigid labels \cite{sensoy2018evidential}. The intuition behind EDL lies in modeling the network predictions as a distribution over the categorical SoftMax outputs. Sampling from this distribution yields the realization of the uncertainty through the variance of the distribution.

The Dempster-Shafer theory of evidence (DST) \cite{shafer1992dempster} and Subjective Logic (SL) provides a well-defined framework for quantifying class uncertainties. The approach suggests assigning a belief mass $b_k$ for each of the $K$ classes of the network. $b_k$ represents the belief that the truth can be any of the possible states. Hence the overall uncertainty $u$ can be expressed by computing the residual sum of all belief masses:
\begin{equation}\label{SL}
u +\sum_{k=1}^{K} b_k = 1
\end{equation}

In EDL, The SoftMax layer is removed, and the neural network is trained using a specific loss function to output evidence vectors $e_k$. These evidence vectors represent the amount of support on how confident a sample is, given a particular class. The vectors relate to the belief masses and uncertainties as follows: 
\begin{equation}
b_k=\frac{e_k}{S}, \quad u=\frac{K}{S}, \quad \textit{where,} \quad S=\sum_{k=1}^{K} (e_k+1)
\label{Eq: uncertainty}
\end{equation}
It is interesting to note that the uncertainty value from Equation \ref{Eq: uncertainty} is inversely proportional to $S$, which represents the sum of evidence and the number of labels. This notes that a zero-evidence vector would yield a belief mass $b_k=0$, and the maximum uncertainty $u=1$. 

The likelihood function for classification models learns over a categorical distribution. EDL places a higher order distribution over the parameters of the categorical distribution (label probabilities) to allow for prediction uncertainty estimation $u$. To ensure a tractable posterior, it is essential to select a prior conjugate distribution, such as the Dirichlet distribution. Parameterized by $\alpha=[\alpha_1,...,\alpha_K ]$, the Dirichlet distribution's density function is expressed by:

\begin{equation}
D(\mathbf{p} \mid \alpha)=\frac{1}{\beta(\alpha)}\prod_{k=1}^{K}p_k^{\alpha_k-1}
\end{equation}
Where $\mathbf{p}$ is the probability mass function, $\beta(\alpha)$ is the Beta function, $p_k$ is the probability for label $k$ and $\alpha_k$ can be calculated as $\alpha_k = e_k +1$. The probability of each class is equivalent to the expectation of the Dirichlet distribution:
\begin{equation}
\mathbb{E}[D(\alpha)]=\frac{\alpha_k}{S}
\end{equation}
The $S$ here is taken from Equation \ref{Eq: uncertainty} and can be referred to as the strength of the Dirichlet distribution.

A Kullback-Leibler (KL) divergence is used to regularize the prediction of the network. The term measures the similarities of the predicted distribution to a uniform Dirichlet distribution with $\alpha_k = 1$ $\forall$ $k \in K$ or zero evidence. The KL term is added to the loss function, and therefore, penalizes samples that deviate from class categories. The KL term is given by: 

\begin{multline}
        KL[D(p_i \mid \tilde \alpha_i) \mid \mid D(p_i \mid 1) ] =\\
        \log \bigg(\frac{\Gamma(\sum_{k=1}^{K}\tilde \alpha_{ik})}{\Gamma(K)\prod_{k=a}^{K}\Gamma(\tilde \alpha_{ik})}\bigg)
         + \sum_{k=1}^{K} (\tilde \alpha_{ik} -1)\Big[\psi(\tilde \alpha_{ik}) - \psi\Big(\sum_{j=1}^{K}\tilde \alpha_{ik}\Big) \Big]
\end{multline}
Where $\tilde \alpha_{ik} = y_i + (1-y_i)\odot \alpha_i$ is the modified Dirichlet parameter. $\Gamma$ is the Gamma function and $\psi$ is the Digamma function.

\subsubsection{Conformal Prediction}
CP, an emerging method for quantifying uncertainty, has garnered attention in the fields of machine learning and computer vision. Renowned for its simple implementation, computational efficiency, and versatility across various models, CP techniques have become increasingly prominent. They can be applied to a wide range of pre-trained models, including decision support models, shallow multi-layer perceptrons, deep neural networks, and more. Additionally, they can be deployed for various tasks such as classification or regression. Furthermore, these methods do not require any assumptions about the underlying data distribution. Additionally, they provide a guarantee of validity, even when working with a finite number of samples. Hence, CP proves to be indispensable across a broad spectrum of real-world applications.

Unlike other uncertainty quantification methods, the output of the CP algorithm is a prediction set of all possible labels such that the true label is included within this set with a user-defined confidence level. The implementation of CP only requires a model $f$ trained conventionally on a given dataset, and an additional small independent and identically distributed (i.i.d) dataset, which we will refer to as the calibration dataset. To understand the underlying details of CP, consider a classification task where a deep learning model is trained on a dataset of images $x_{\text{train}}$ to correctly classify each input image to its corresponding $y_{\text{train}}$ label chosen from $K$ total classes. The class probability outputs, i.e., SoftMax function $\sigma$, are utilized as a notion of model uncertainty and used to derive the conformity scores as $S =\sigma(f(x))$. Furthermore, a designated $C$ subset of sample images and their true labels are reserved for the calibration, where $C = \{(X_n, Y_n), n=1,\ldots,N\}$. Both $S$ and $C$ are used to construct a function $\mathcal{T}$ that returns a prediction set of possible labels. The process of finding $T$ includes passing all samples $X_n$ $\in$ $C$ to the model, collecting their $N$ conformity score of the correct label $S(X_n, Y_n) = \sigma(f(X_n)_y$, and finally calculating $\hat{q}$ which is the empirical quantile of the scores using the user-defined error rate $\alpha \in [0,1]$ and the size of the calibration set $N$:
\begin{equation}
   \hat{q} = \frac{\left\lceil(1+N)(1-\alpha)\right\rceil}{N} 
   \label{Eq: quantile}
\end{equation}

For a new fresh pair $(x_{\text{test}},y_{\text{test}})$, the conformity score is calculated, and the prediction set is generated to include the labels whose score values are less than $\hat{q}$:
\begin{equation}
\mathcal{T}(x_{\text{test}}) = \{y: S(x_{\text{test}}) \leq \hat{q}\}
\end{equation}

The resulting prediction set guarantees the conformal coverage criteria, with the following probability:
\begin{equation}
1-\alpha \leq P(Y_{\text{test}} \in T(x_{\text{test}})) \leq (1-\alpha)+\frac{1}{1+M}
\end{equation}

The design of the scoring function is a critical element that requires careful engineering. This scoring function encapsulates all the information embedded in the model; thus, a robust scoring function is essential for generating a meaningful conformal set. It's important to emphasize that the preceding steps can be extended to include a variety of tasks, including regression. The sole variation in the steps outlined above lies in the choice of a scoring function tailored to the specific requirements of the given task.

\begin{figure}[h]
    \centering
    \includegraphics[width=0.7\linewidth]{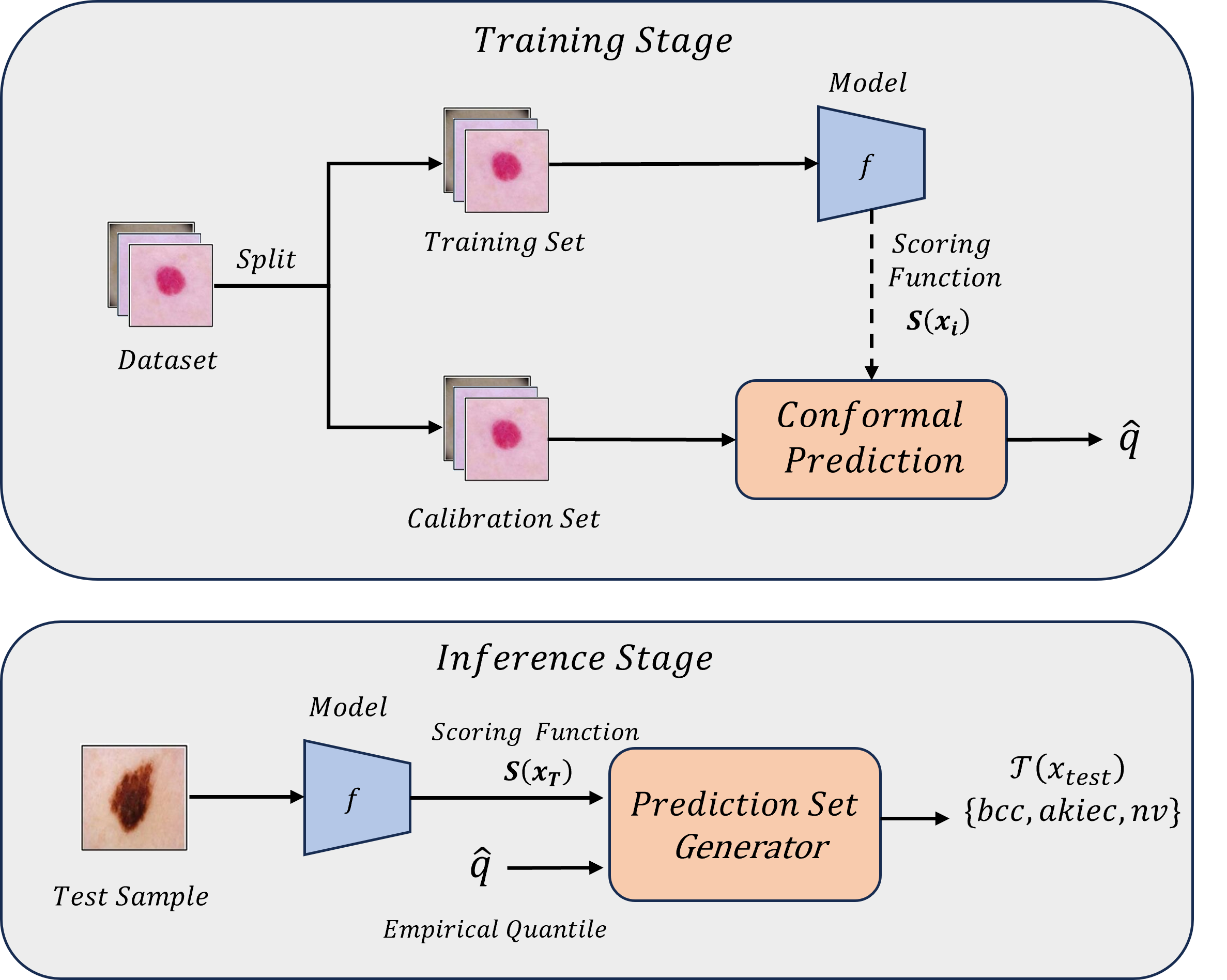} 
    \caption{Conformal Prediction Framework. The dataset is split into a training set, a calibration set, and a testing set. Conformal Prediction utilizes the scoring function and calibration set to generate a quantile value $\hat{q}$. At inference, a test sample is passed to the train network, and the output score along with the quantile value is used to generate prediction sets.}
    \label{fig: ICP}
\end{figure}


\subsection{Uncertainty Quantification Settings}

As previously indicated, ResNet-18 serves as the foundational architecture for implementing uncertainty quantification methods. However, distinct modifications are necessitated to incorporate each of the three uncertainty quantification techniques. In this section, we elaborate on the required design settings tailored for the implementation of each method. Furthermore, we address any supplementary training configurations and methodologies employed in the integration of the core model to the uncertainty quantification technique. It's crucial to note that due to the differing implementation procedures required by each uncertainty quantification technique—including distinct training approaches and loss functions—each model will ultimately exhibit varying classification accuracies, subject to the training of the networks.

CP, as a post-hoc technique, obviates the need for retraining the core network. The implementation simply involves reserving a portion of the dataset, unused during the training phase, for calibration. Following the work in \cite{angelopoulos2021gentle}
, we assigned 1000 calibration data samples. Furthermore, we have assigned the confidence level to 90\%. The overall algorithm is summarized in Algorithm \ref{alg:1}. Moreover, the effect of changing the calibration size and the confidence level are further studied in Section \ref{ssec: 3.4}. 

\begin{algorithm}[H]
  \caption{APS}\label{alg:1}
  \begin{algorithmic}[1]
    \State \textbf{input}: (model, calibration set, new input)
    \State \textbf{get scores:} Apply the scoring function to all training images as:
      $
        E_j = \sum_{i=1}^{k'} (\hat{\pi}_{(i)}(x_j)
      $
      where $k'$ is the model's ranking of the true class $y_j$, and $\hat{\pi}_{(i)}(x_j)$ is $i^{th}$ largest score for the $j^{th}$ input.
    \State \textbf{Compute $\hat{q}$:} The value corresponding to the $1-\alpha$ quantile of $E_j$
    \State \textbf{Output prediction set:} $\sum_{i=1}^{k^*}(\hat{\pi}_{(i)}(x_{n+1})) \le \hat{q}$
  \end{algorithmic}
\end{algorithm}

For the implementation of MCD, a dropout layer with a rate of 0.5 was added to the primary ResNet-18 model, positioned before the last classification layer. The uncertainty estimation process entailed conducting 1000 forward passes during the inference stage.

Unlike CP and MCDropout, the EDL algorithm requires retraining the model with different settings. The conventional cross-entropy loss function is replaced with a mean square error loss function, tailored to estimate the $\alpha$ parameters of the Dirichlet distribution, which is used to estimate the predictive uncertainty. The loss function is presented by: 
\begin{equation}
    \mathcal{L}_i(\theta)=\sum_{j=1}^{K}(y_{ij}-\frac{\alpha_{ij}}{S_i})^2+ \frac{\alpha_{ij}(S_i-\alpha_{ij})}{S_i^2(S_i+1)}
\end{equation}
where $y_{ij}$ is the ground truth class label, $\alpha_{ij}$ is the corresponding Dirichlet parameter, and $S$ is the strength of the Dirichlet distribution, as expressed in Equation \ref{Eq: uncertainty}. Similar to the previous optimizer settings, the Adam optimizer is used with a learning rate of 1e-5, and the model was trained for 150 epochs. 

\section{Results}
\label{sec:3}

\subsection{Evaluation of Uncertainty Quantification methods} \label{ssec:3.3}
This section presents a quantitative assessment and comparative analysis of the three uncertainty quantification methods: CP, MCD, and EDL. In this study, our investigation relies on examining the distribution of quantified uncertainty values assigned by each approach. To assess performance comprehensively, we utilized both the mean uncertainty value of correctly classified samples and that of wrongly classified samples. A consistent and robust uncertainty quantification approach lies in its ability to assign higher uncertainty values to wrongly classified samples compared to correctly classified samples.

Initially, the base classification model was trained without the incorporation of any uncertainty method. The model achieves an 86.7\% accuracy rate on the HAM10000 testing dataset that contains 1103 samples. In the context of CP, we set a user-defined confidence interval of $1-\alpha$, where $\alpha$ is chosen to be 0.10, corresponding to a 90\% confidence level. This implies that the ground truth label is encompassed within the generated prediction sets with a 90\% confidence bound. A larger prediction set indicates higher predictive uncertainty, while a smaller set implies lower uncertainty. As mentioned earlier, CP operates as an add-on algorithm, acting on the SoftMax probabilities produced by the core model. Consequently, deploying CP does not alter the initial accuracy of the model. In the context of predictive uncertainty, the average uncertainty value of all correctly classified samples is 0.4, while the average uncertainty value of all wrongly classified samples is 0.79.

Meanwhile, for the MCD method, 1000 forward passes are executed during inference to estimate the model's uncertainty, resulting in accuracies of 85.8\%. Noteworthy is that correctly classified examples in MCD are associated with a mean uncertainty value of 0.01, whereas wrongly classified examples exhibit a mean uncertainty value of 0.09. In contrast, EDL takes a distinct approach by being trained from scratch using the mean square error loss that explicitly accounts for uncertainties. The EDL model achieves an accuracy of 85.7\%, with a mean uncertainty value of 0.19 assigned to correctly classified samples and 0.51 to wrongly classified examples. 

\begin{figure}[h]
    \centering
    \begin{subfigure}{0.72\linewidth}
        \includegraphics[width=\linewidth]{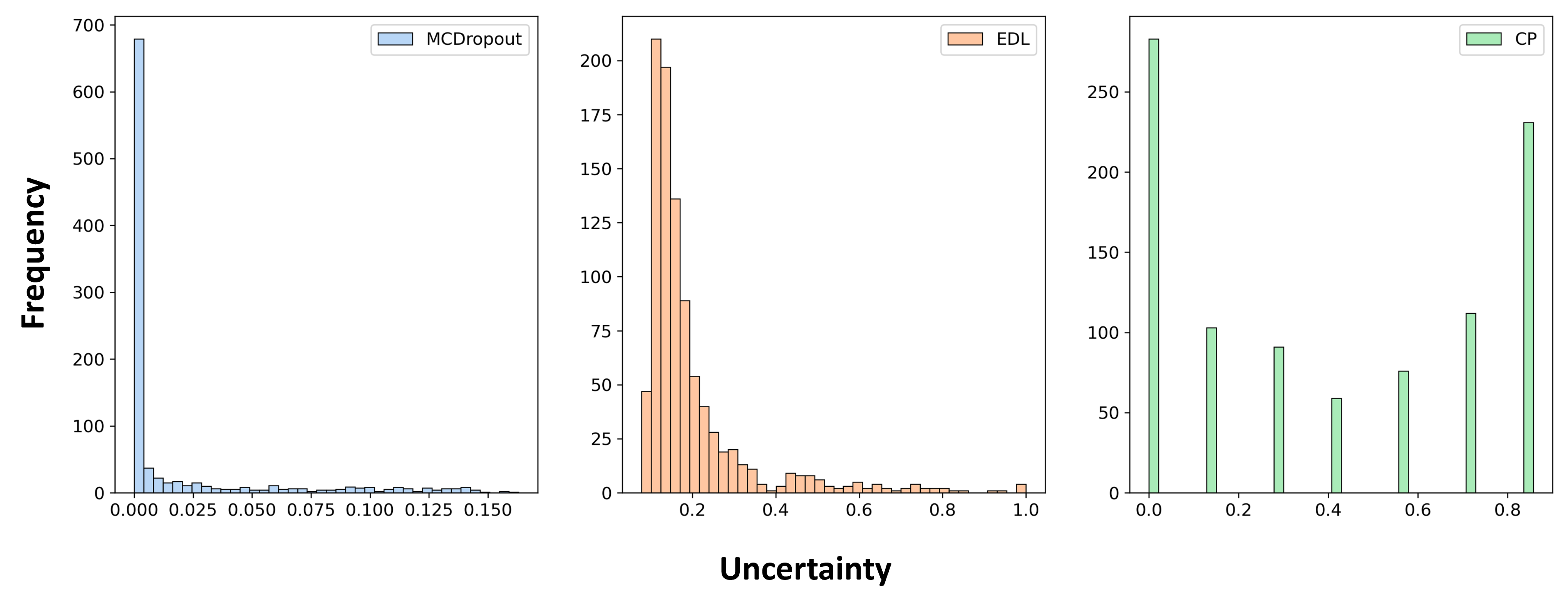} 
        \caption{Correctly classified samples}
        \label{fig:subfigure_a}
    \end{subfigure}
    \hfill
    \begin{subfigure}{0.7\linewidth}
        \includegraphics[width=\linewidth]{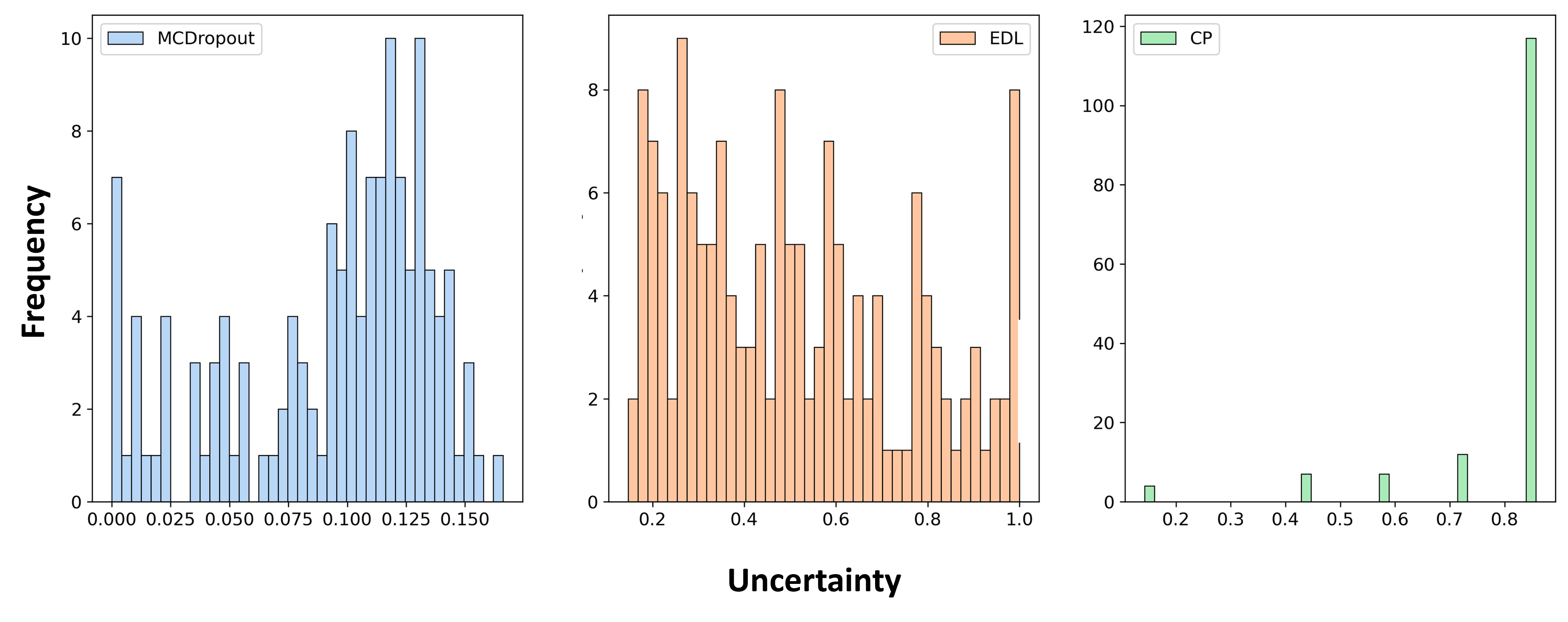}
        \caption{Wrongly classified samples}
        \label{fig:subfigure_b}
    \end{subfigure}
    \caption{The histogram distribution of uncertainty values assigned by each uncertainty quantification algorithm to both correctly and wrongly classified samples. Results illustrate that while both MCD and EDL assign diverse values of uncertainties to wrongly classified samples, CP, on the other hand, assigns high uncertainty values to those samples.}
    \label{fig:UQhist_Resnet18}
\end{figure}

Additionally, we have conducted a similar analysis of the three uncertainty quantification methods on the DMF dataset and the BCM dataset. The detailed results are shown in Table \ref{tab:compareUQ_models}. Moreover, the histogram distributions of the assigned uncertainty values for correctly and wrongly classified samples, across the three aforementioned uncertainty quantification methods are shown in Figure \ref{fig:UQhist_Resnet18}.


\begin{table}[h]
    \centering
    \begin{tabular}{|c|c|c|c|}
        \hline
        \textbf{UQ method} & \textbf{Accuracy} & $\mathbf{U_{\text{correct}} \pm \text{std}}$ & $\mathbf{U_{\text{wrong}} \pm \text{std}}$ \\
        \hline
        \multicolumn{4}{|c|}{\textbf{HAM10000}} \\
        \hline
        MCD & 85.8\% & $0.01 \pm 0.03$ & $0.09 \pm 0.04$ \\
        \hline
        EDL & 85.7\% & $0.19 \pm 0.13$ & $0.51 \pm 0.25$ \\
        \hline
        CP & 86.7\% & $0.40 \pm 0.30$ & $0.79 \pm 0.15$ \\
        \hline
        \multicolumn{4}{|c|}{\textbf{DMF}} \\
        \hline
        MCD & 75.2\% &$0.05 \pm 0.03$ & $0.08 \pm 0.02$ \\
        \hline
        EDL & 70.4\% &$0.24 \pm 0.17$ & $0.47 \pm 0.11$ \\
        \hline
        CP & 72.7\% &$0.33 \pm 0.25$ & $0.57 \pm 0.18$ \\
        \hline
        \multicolumn{4}{|c|}{\textbf{BCM}} \\
        \hline
        MCD & 98.0\% & $0.01 \pm 0.02$ & $0.07 \pm 0.03$ \\
        \hline
        EDL & 98.2\% & $0.42 \pm 0.06$ & $0.48 \pm 0.06$ \\
        \hline
        CP & 98.6\% & $0.34 \pm 0.29$ & $0.78 \pm 0.16$ \\
        \hline
    \end{tabular}
    \caption{Comparison of UQ Algorithms: Accuracy, Average Uncertainty Estimates for HAM10000, DMF, and BCM datasets. The table presents model accuracy along with average uncertainty values for both correctly classified samples and wrongly classified samples, as determined by the core network and per their respective ground truth labels. Algorithms include Monte Carlo Dropout (MCD), Evidential Deep Learning (EDL), and Conformal Prediction (CP).}
    \label{tab:compareUQ_models}
\end{table}

\begin{figure}[h!]
    \centering
    \begin{subfigure}{0.80\linewidth}
        \includegraphics[width=\linewidth]{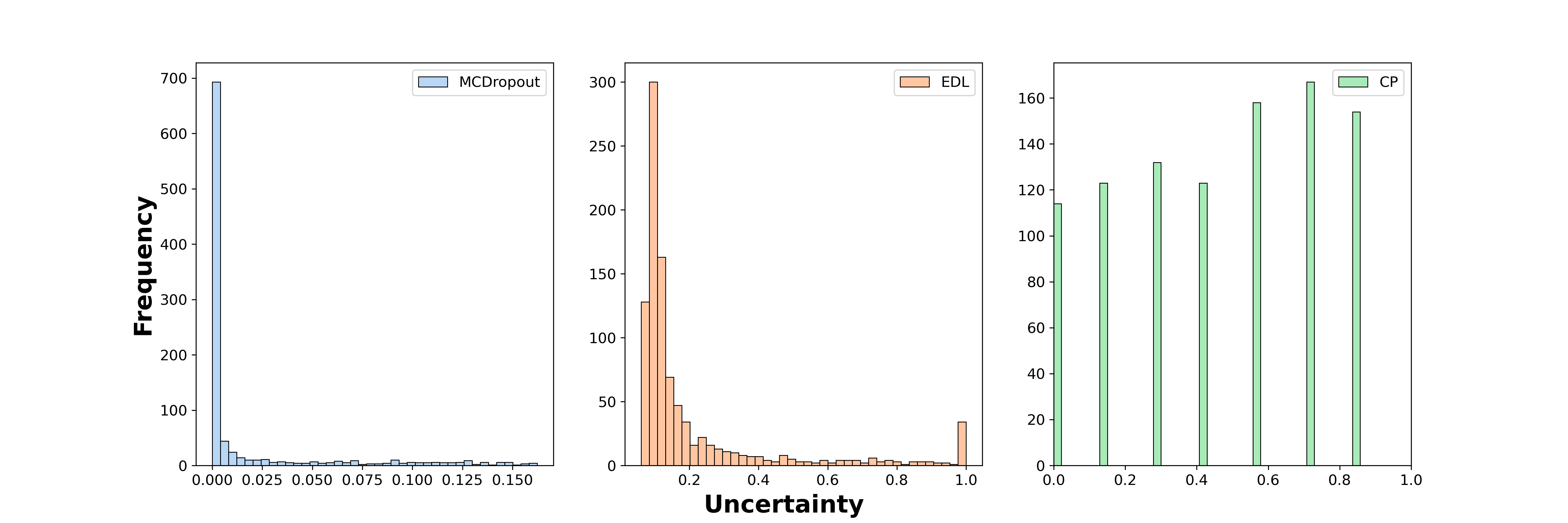} 
        \caption{Correctly classified samples: ResNet-50}
        \label{fig:subfigure_a}
    \end{subfigure}
    \hfill
    \begin{subfigure}{0.80\linewidth}
        \includegraphics[width=\linewidth]{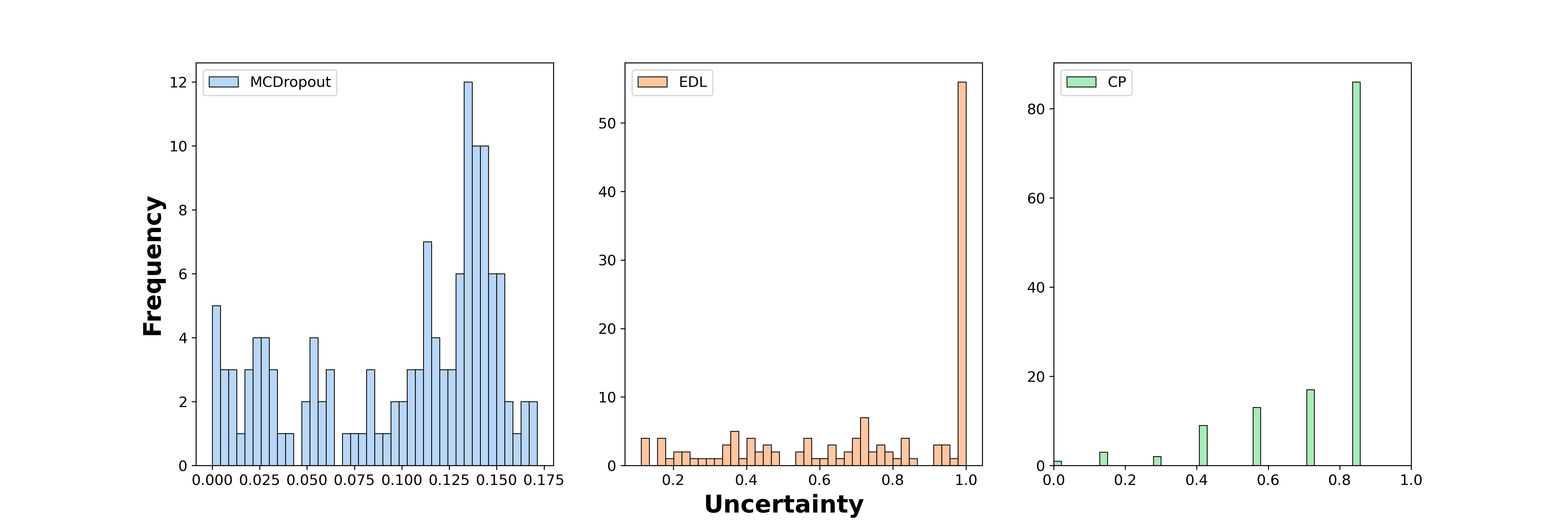}
        \caption{Wrongly classified samples: ResNet-50}
        \label{fig:subfigure_b}
    \end{subfigure}
    
    \begin{subfigure}{0.80\linewidth}
        \includegraphics[width=\linewidth]{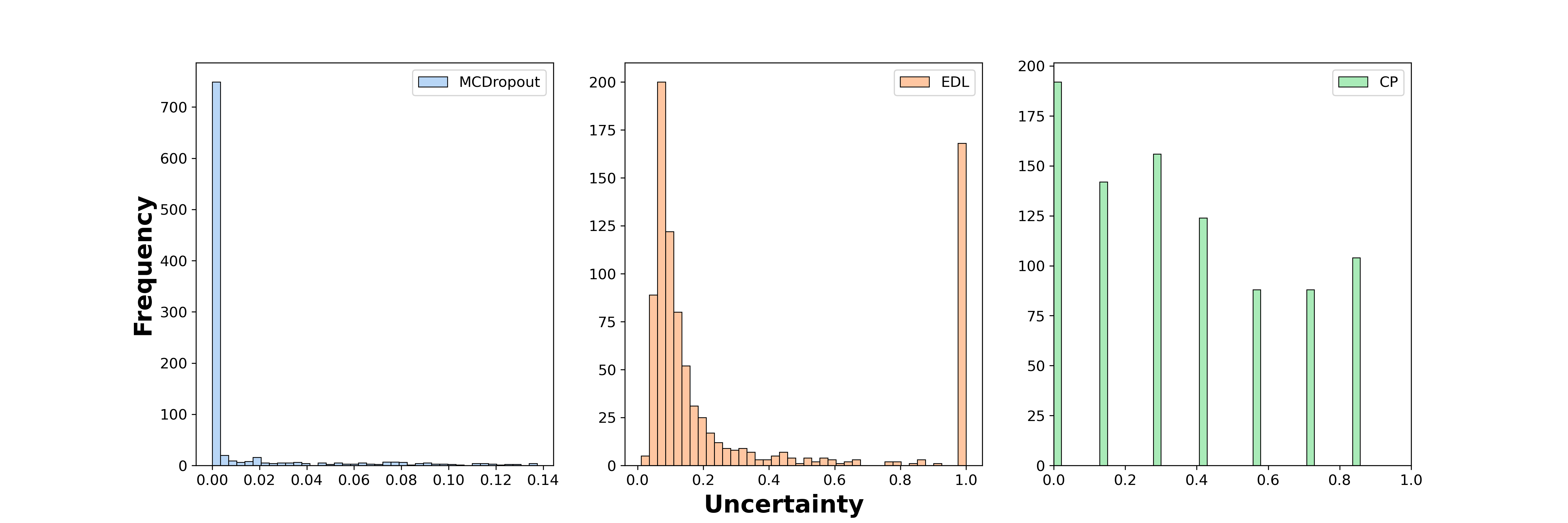} 
        \caption{Correctly classified samples: VGG-11}
        \label{fig:subfigure_c}
    \end{subfigure}
    \hfill
    \begin{subfigure}{0.80\linewidth}
        \includegraphics[width=\linewidth]{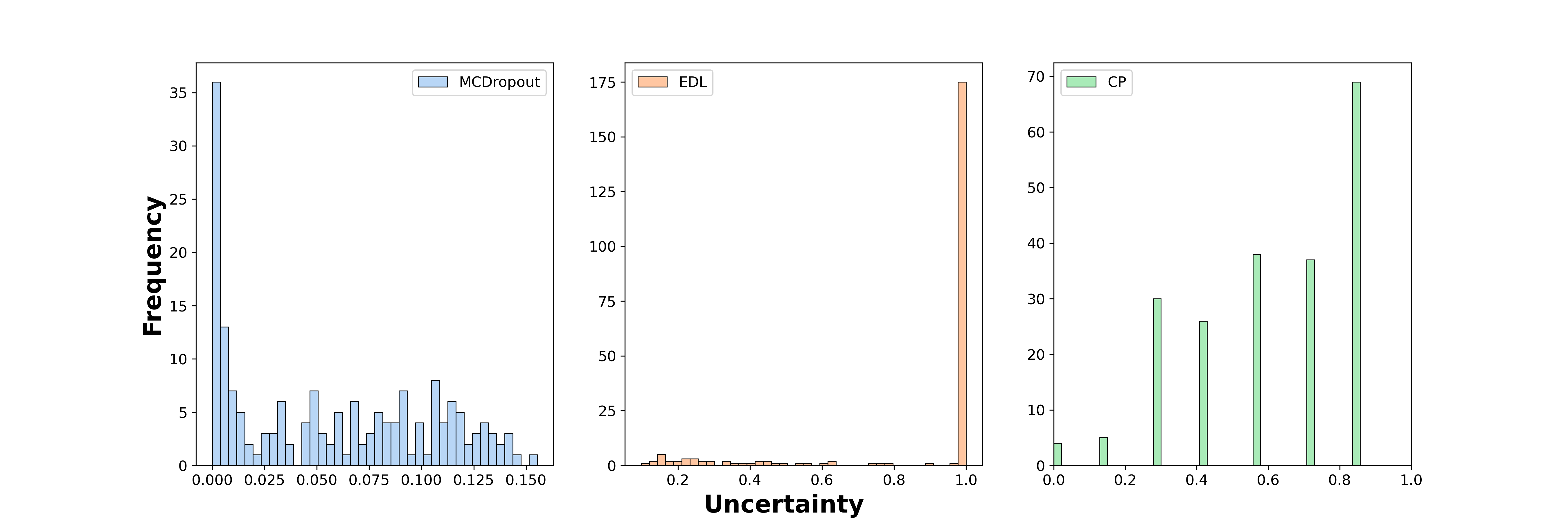}
        \caption{Wrongly classified samples: VGG-11}
        \label{fig:subfigure_d}
    \end{subfigure}
    
    \caption{Histogram of uncertainty values assigned to correctly classified samples and wrongly classified samples using 2 different deep learning models: ResNet-50 and VGG-11. The three uncertainty quantification methods are MCDropout, EDL, and CP}
    \label{fig:UQhist_ReNetVGG}
\end{figure}

Next in our study, we extended our analysis and experimental investigations by including two additional models: ResNet-50 and VGG-11. The objective is to evaluate how altering the core network impacts the uncertainty results generated by each of the employed methods. The structure of ResNet-50 shares similarities with ResNet-18, albeit being deeper with over 50 convolutional layers. VGG-11 on the other hand, consists of multiple layers of convolutional and max-pooling operations, followed by fully connected layers. Despite its straightforward design, VGG-11 has demonstrated competitive performance across a wide range of image classification tasks. The same training procedures that were implemented in Section \ref{Sec: model specs} are followed in these experiments. The ResNet-50 model achieved similar results to the initial experiments. We observe that there is no significant impact on uncertainty values generated by using ResNet-50. In contrast, VGG-11 network did not achieve high classification accuracies, and hence the worse performance of the CP uncertainty quantification. We record the accuracies and average uncertainties assigned by each method in Table \ref{tab:compareUQ_ResVGG}. Moreover, the distribution of the uncertainty values assigned for both the correctly classified examples and the wrongly classified examples for the three uncertainty methods are shown in Figure \ref{fig:UQhist_ReNetVGG}

\begin{table}[h]
    \centering
    \begin{tabular}{|c|c|c|c|c|c|c|}
        \hline
        \multicolumn{1}{|c|}{\multirow{2}{*}{\textbf{UQ algorithm}}} & \multicolumn{3}{c|}{\textbf{ResNet-50}} & \multicolumn{3}{c|}{\textbf{VGG-11}} \\
        \cline{2-7}
         & \textbf{Accuracy} & $\mathbf{U_{\text{correct}} \pm \text{std}}$ &
        $\mathbf{U_{\text{wrong}} \pm \text{std}}$ & \textbf{Accuracy} & $\mathbf{U_{\text{correct}} \pm \text{std}}$ &
        $\mathbf{U_{\text{wrong}} \pm \text{std}}$ \\
        \hline
        MCD & 88.1\% & $0.01 \pm 0.03$ & $0.09 \pm 0.05$ & 83.7\% & $0.01 \pm 0.02$ & $0.05 \pm 0.04$ \\
        \hline
        EDL & 87.4\% & $0.21 \pm 0.22$ & $0.73 \pm 0.29$ & 80.2\% & $0.30 \pm 0.36$ & $0.87 \pm 0.27$ \\
        \hline
        CP & 88.0\% & $0.46 \pm 0.28$ & $0.75 \pm 0.18$ & 81.0\% & $0.35 \pm 0.28$ & $0.61 \pm 0.23$ \\
        \hline
    \end{tabular}
    \caption{Comparison of UQ Algorithms: Accuracy, average Uncertainty Estimates for ResNet-50 and VGG-11. The table presents model accuracy along with average uncertainty values for both correctly classified samples and wrongly classified samples, as determined by the core network and per their respective ground truth labels. Algorithms include Monte Carlo Dropout (MCD), Evidential Deep Learning (EDL), and Conformal Prediction (CP).}
    \label{tab:compareUQ_ResVGG}
\end{table}

\begin{figure}[htbp]
    \centering
    \includegraphics[width=0.8\textwidth]{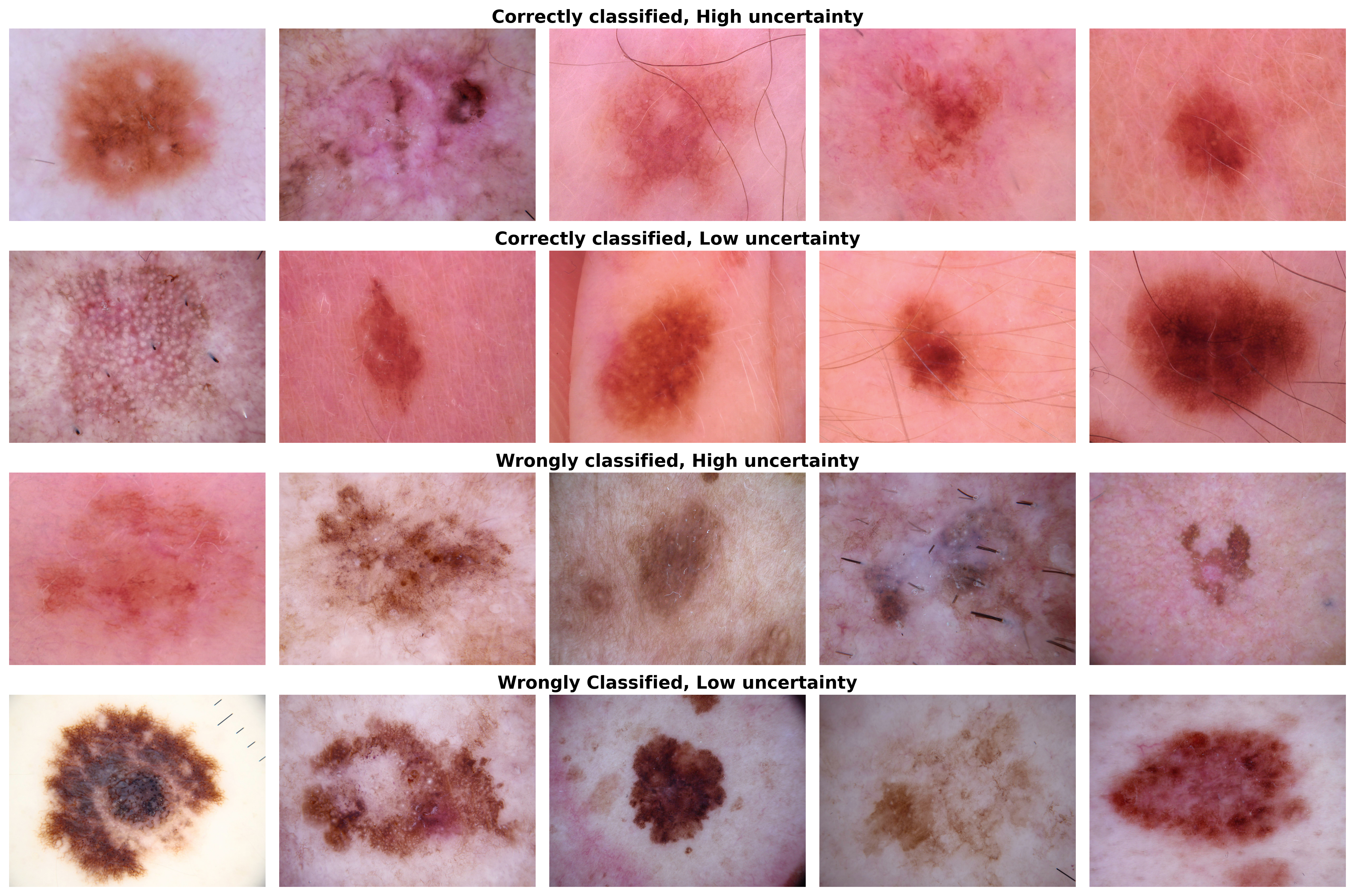}
    \captionsetup{justification=justified,singlelinecheck=false}
    \caption{Analysis of samples with corresponding uncertainty values. Samples are categorized into correctly classified and wrongly classified groups, with their uncertainty quantified using the conformal prediction algorithm. Misclassified samples often exhibit a noticeably different skin tone compared to correctly classified ones. Additionally, misclassified samples with low uncertainty display more visible skin lesion patches compared to those with high uncertainty, possibly influencing the model's predictive confidence.}
    \label{fig:analysis}
\end{figure}

Finally, we conducted further analysis by examining samples from the HAM10000 dataset along with their corresponding uncertainty values, categorized into distinct groups based on their classification outcomes by the core network. These groups were further stratified into correctly classified and wrongly classified samples, with their uncertainty levels quantified using the conformal prediction algorithm. The samples are shown in Figure \ref{fig:analysis}. Notably, a discernible difference in skin tone is observed between misclassified and correctly classified samples. Such potential bias in the dataset can adversely affect network accuracy. Furthermore, we observed that misclassified samples with low uncertainty exhibited more prominent skin lesion patches compared to those with high uncertainty. This disparity in patch visibility could account for the lower uncertainty associated with the former group. 

\subsection{Sensitivity of Conformal Prediction Parameters} \label{ssec: 3.4}
In this section, we delve into the details of the model's performance by conducting a comprehensive sensitivity analysis on key CP parameters. We investigate the impact of three pivotal factors: first, the selection of the scoring function; second, the adjustment of the confidence level $\alpha$; and third, the variation in the calibration set size. Through this exploration, we aim to study the dependencies between these parameters and the model's performance in the context of uncertainty estimation.
\subsubsection{Scoring Function}
The scoring function stands out as a critically significant element within the context of CP. While the CP algorithm consistently ensures marginal coverage by regulating the size of the prediction set, 
it is the scoring function that ultimately determines the utility of the set's content
the scoring function distinguishes whether the predictions included in the set are deemed useful or not. To this end, we study the effect of altering the scoring function and present analysis on two key terms: 1) The distribution of uncertainties on correctly classified samples Vs. wrongly classified samples and 2) the size of the prediction set. 

To determine the effect of the scoring function, an additional variation of CP, known as Regularized Adaptive Prediction Sets (RAPS) \cite{angelopoulos2020uncertainty} is implemented. RAPS takes the SoftMax outputs, similar to the APS method implemented in our experimental results, however, it adds a regularization term to the extremely large prediction sets to discourage their presence in the final prediction set. As a result, the average prediction set size produced by RAPS is much smaller than that produced by APS. The RAPS algorithm is summarized in Algorithm \ref{alg:2}
\begin{algorithm}[H]
  \caption{RAPS}\label{alg:2}
  \begin{algorithmic}[1]
    \State \textbf{input}: (model, calibration set, new input)
    \State \textbf{get scores:} Apply the scoring function to all training images as:
      $
        E_j = \sum_{i=1}^{k'} (\hat{\pi}_{(i)}(x_j) + \lambda \textbf{1}[i> k_{\text{reg}}]),
      $
      where $k'$ is the model's ranking of the true class $y_j$, and $\hat{\pi}_{(i)}(x_j)$ is $i^{th}$ largest score for the $j^{th}$ input.
    \State \textbf{Compute $\hat{q}$:} The value corresponding to the $1-\alpha$ quantile of $E_j$
    \State \textbf{Output prediction set:} $\sum_{i=1}^{k^*}(\hat{\pi}_{(i)}(x_{n+1}) +\lambda \textbf{1}[j> k_{\text{reg}}]) \le \hat{q}$
  \end{algorithmic}
\end{algorithm}

\begin{figure}[h]
    \centering
    \begin{subfigure}{0.8\linewidth}
        \includegraphics[width=\linewidth]{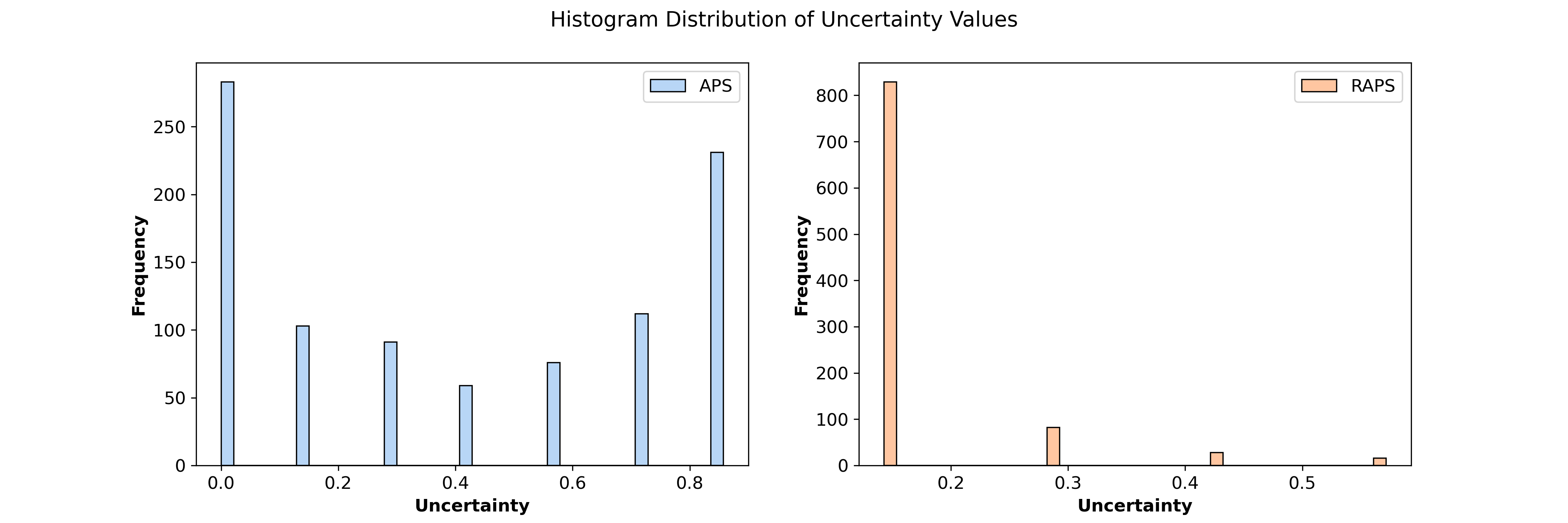} 
        \caption{Correctly classified samples}
        \label{fig:subfigure_a_RAPS}
    \end{subfigure}
    \hfill
    \begin{subfigure}{0.8\linewidth}
        \includegraphics[width=\linewidth]{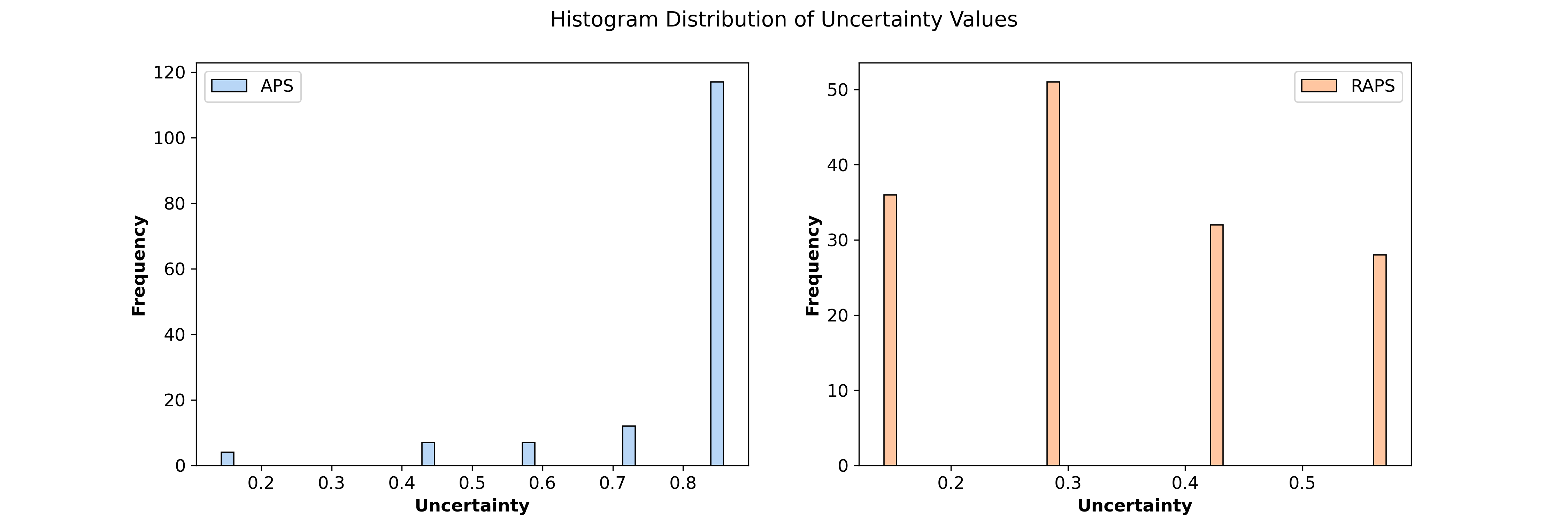}
        \caption{Wrongly classified samples}
        \label{fig:subfigure_b_RAPS}
    \end{subfigure}
    \caption{Comparing the effect of APS Vs. RAPS on uncertainty assignments for both correctly classified samples and misclassified samples. The figure illustrates that smaller prediction set sizes produced by RAPS, often indicating lower uncertainty, are generally preferred. However, challenges arise in cases of misclassified samples where larger set sizes, indicating increased uncertainty, may incur penalties from the regularization term}
    \label{fig:CPhist}
\end{figure}

Results presented in Table \ref{Tab: T2} show that RAPS produces an average prediction set size smaller than that generated by APS. While smaller prediction set sizes are typically preferred as they imply greater certainty, this characteristic may pose challenges, particularly in instances of misclassified samples where the regularization term penalizes larger set sizes that show more uncertainty, as shown in Figure \ref{fig:CPhist}

\begin{table}[h]
    \centering
    \begin{tabular}{|c|c|c|c|}
        \hline
        \textbf{Conformal Method} & \textbf{$C_{correct}$} & \textbf{$C_{wrong}$} & \textbf{Average}\\
        \hline
        APS & 2.84 & 5.54 & 4.19 \\
        \hline
        RAPS & 1.19 & 2.35 & 1.77  \\
        \hline
    \end{tabular}
    \caption{Comparison between the average prediction set size produced by APS and RAPS algorithms for correctly classified samples and misclassified samples. Results illustrate that RAPS is capable of generating a smaller average set size compared to APS}
    \label{Tab: T2}
\end{table}

\subsubsection{Confidence Level}

\begin{figure}[h]
    \centering \includegraphics[width=0.5\linewidth]{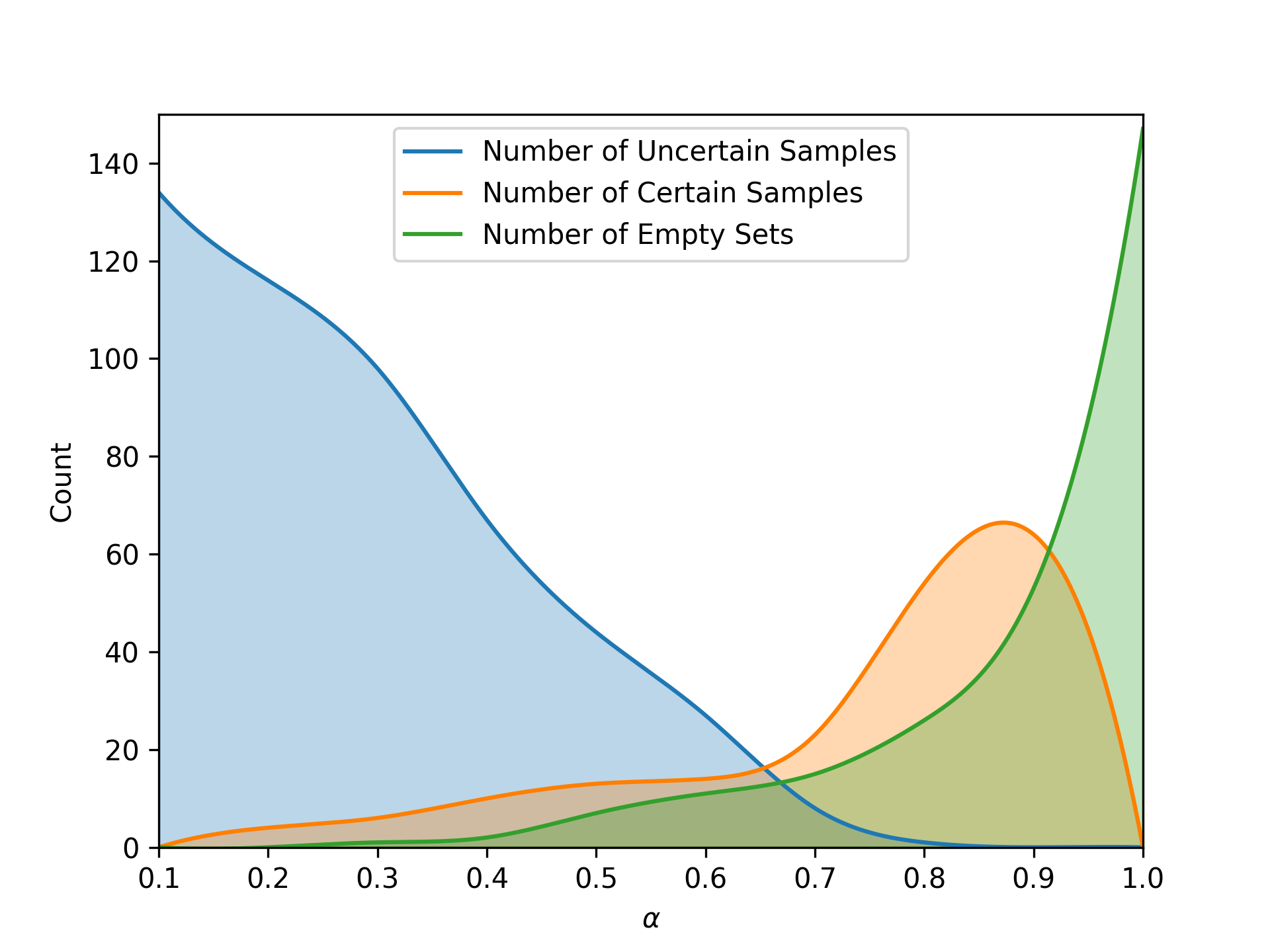} 
    \caption{The relationship of varying the value of $\alpha$ that corresponds to the confidence level and the count of uncertain samples, certain samples, and empty sets. The figure illustrates that increasing the value of $\alpha$ results in producing more empty sets ($\alpha$ =1).}
    \label{fig: alpha}
\end{figure}

The confidence level ($1-\alpha$) is a user-defined parameter, representing the probability of predicting the true label within the generated prediction set. The value of $\alpha$ is also used to compute the empirical quantile that splits the output scores into conforming and nonconforming scores. According to Equation \ref{Eq: quantile}, increasing $\alpha$ reduces the value of the quantile. at $\alpha =1.0$, the value of the quantile is $0$, hence the prediction set will produce empty sets of predictions. To further investigate the impact of increasing the value of $\alpha$, we examined the algorithm's uncertainty assignment through the samples of misclassified data. As illustrated in Figure \ref{fig: alpha}, it is evident that CP assigns high uncertainty values to a higher count of samples at $\alpha =0.1$. This is an expected behavior as the samples are wrongly classified, and hence they belong to the uncertain category. As the value of 
$\alpha$ increases and the quantile value decreases, leading to a reduction in the number of samples generated with high uncertainty. Consequently, the count of certain samples increases, and eventually, those samples turn into empty prediction sets at higher $\alpha$ values.  

\subsubsection{Calibration set size}

\begin{figure}[h]
    \centering \includegraphics[width=0.5\linewidth]{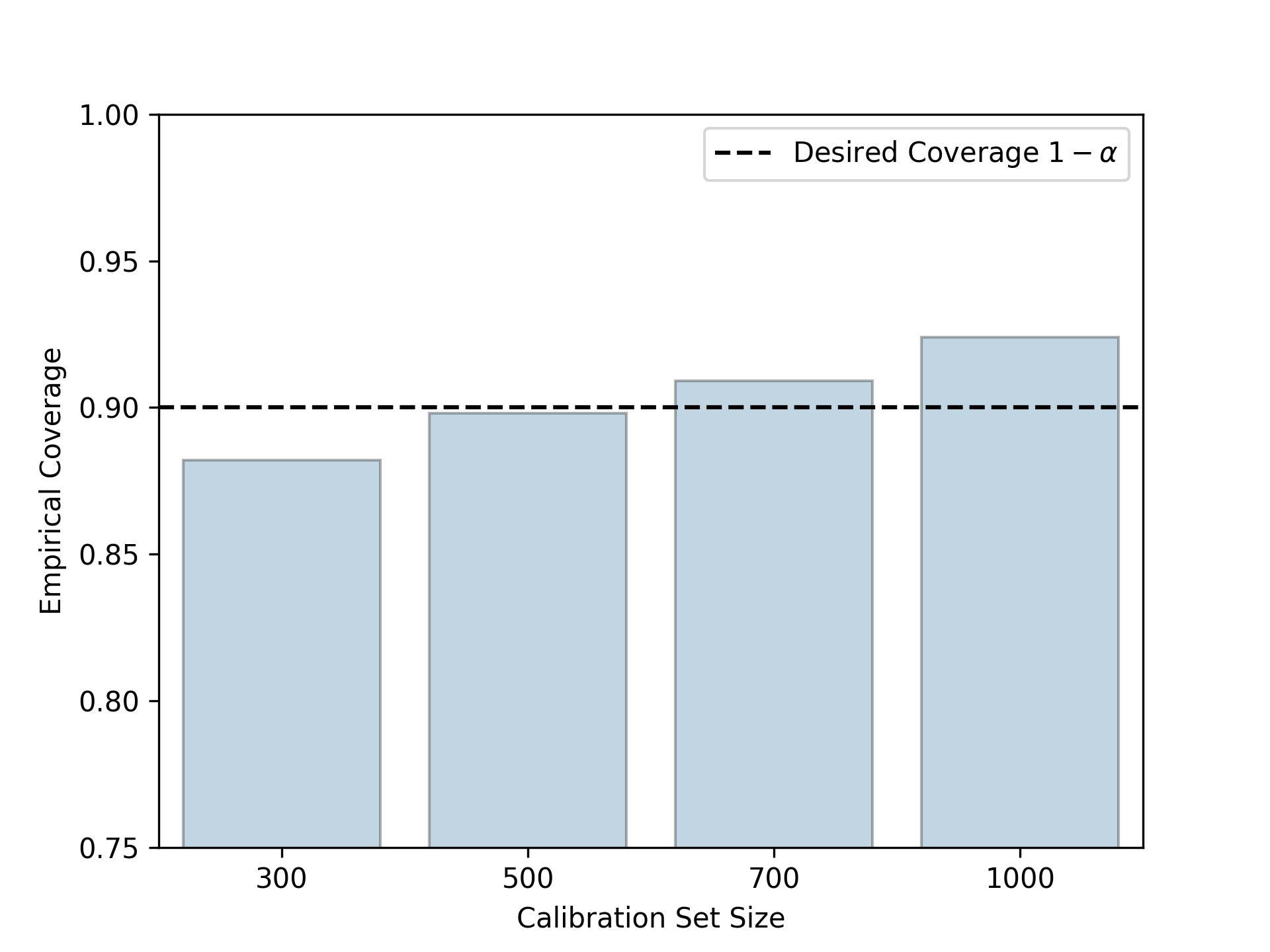} 
    \caption{The relationship between the size of the calibration set and the empirical coverage of conformal prediction algorithm. As illustrated in the Figure, It is advisable to incorporate a large number of samples in the calibration data to mitigate fluctuations in coverage and ensure consistent performance.}
    \label{fig: CSS}
\end{figure}

Varying the calibration set size has an immediate effect on the desired coverage. While conformal prediction guarantees coverage on an average runs on a calibration set, even if the number of samples used in the calibration data is small ~\cite{angelopoulos2021gentle}, It is however recommended to use a high number of samples in the calibration data to overcome any fluctuation in coverage and achieve stable performance. The effect of varying the calibration set size is illustrated in Figure \ref{fig: CSS}, where it can be concluded that increasing the number of samples in the calibration size improves the empirical coverage achieved by the CP algorithm.  

\section{Discussion}

In light of the conducted experiments, the obtained results provide valuable insights and analysis into uncertainty prediction techniques. In high-stakes domains like clinical decision-making, where the ramifications of misclassification can be significant, it is imperative to accompany deep learning algorithms' classification outputs with confidence intervals. Our findings underscore the importance of employing robust uncertainty quantification methods, which yield valuable insights crucial for informed decision-making. Particularly in instances of high uncertainty outputs, involving human expertise in the evaluation process becomes indispensable, augmenting the algorithm's assessment with human judgment and expertise. 

Before discussing the advantages and drawbacks of each technique, it is imperative to emphasize that predictive uncertainty should not be misconstrued as a definitive indicator of classification output. Instead, it serves as supplementary information that enriches the model's robustness. It's typical to find correctly classified samples in the dataset with a mix of both high and low uncertainty values. The variability in uncertainty values is a result of the differences in difficulty levels among individual samples. Ideally, the uncertainty value of the wrongly classified samples should be consistently higher, indicating less confidence in the network's output. As observed in Figure \ref{fig:UQhist_Resnet18}, CP demonstrates a notable trend wherein a significant proportion of misclassified samples are attributed high uncertainty values. Conversely, upon analyzing the outcomes produced by MCD and EDL, we note a more uniform distribution of uncertainty values for misclassified samples. This observation hints at a broader spectrum of uncertainty values, indicating a higher degree of variability in the predictive confidence levels generated by these algorithms.

Compared with MCD and EDL, CP stands out for its simplicity in implementation. The algorithm operates seamlessly, eliminating the need for retraining procedures or modifications to the model's architecture. This crucial feature facilitates the quantification of uncertainties and ensures that the accuracy of the original model remains unaffected. CP, however, requires the calibration dataset to be independent and identical to the training data. This assumption limits the operation of CP in real-world applications, where the tight assumption often does not hold. In MCD, little modifications to the model architecture might be necessary to include dropout layers if not present. Nevertheless, MCD requires running multiple forward passes, contributing to an increased computational load. Moreover, it is crucial to note that high dropout rates have the potential to compromise accuracy, emphasizing the importance of carefully tuning dropout parameters to strike a balance between uncertainty quantification and model performance. Additionally, some studies show that MCD tends to output ill-calibrated uncertainty values that can be unreliable for uncertainty quantification \cite{he2023survey}. EDL, on the other hand, generates sharp uncertainty distribution for the correctly classified samples, but a broad distribution with wrongly classified samples, making it less reliable. The implementation of EDL is relatively more complex when compared to CP and MCD. It requires modifying the core model, changing the training loss function, and retraining the modified model from scratch. Table \ref{tab:uq_comparison} lists a summarization of the advantages and limitations of each of the discussed uncertainty quantification methods.

\begin{table}[htbp]
    \centering
    \caption{Comparison of the advantages and limitations of the uncertainty quantification methods}
    \label{tab:uq_comparison}
    \begin{tabular}{|c|c|c|}
        \hline
        \textbf{UQ Method} & \textbf{Advantages} & \textbf{Limitations} \\
        \hline
        EDL & End-to-end training & Architecture adjustments; Sensitivity to hyperparameters \\
        \hline
        MCD & Simple Implementation & High Computational Overhead \\
        \hline
        CP & Distribution free; Simple Implementation &Exchangability assumption of calibration data \\
        \hline
    \end{tabular}
\end{table}

\subsection{Uncertainty Quantification with Out-of-Distribution Data}

To assess the performance of the three uncertainty quantification methods, Out-of-Distribution (OOD) data is used for testing. OOD is a term that is generally used to refer to the distribution mismatch between the source domain (training data) and the target domain (inferencing data) \cite{fayyad2023exploiting}. Generally, there are two categories of distributional shifts categorized based on the source of discrepancy: semantic shift, where the labels of the target data are different than the source data, and covariate shift, where shifts occur in the input domain, but labels remain unchanged. 

\begin{figure}[h]
    \centering \includegraphics[width=0.6\linewidth]{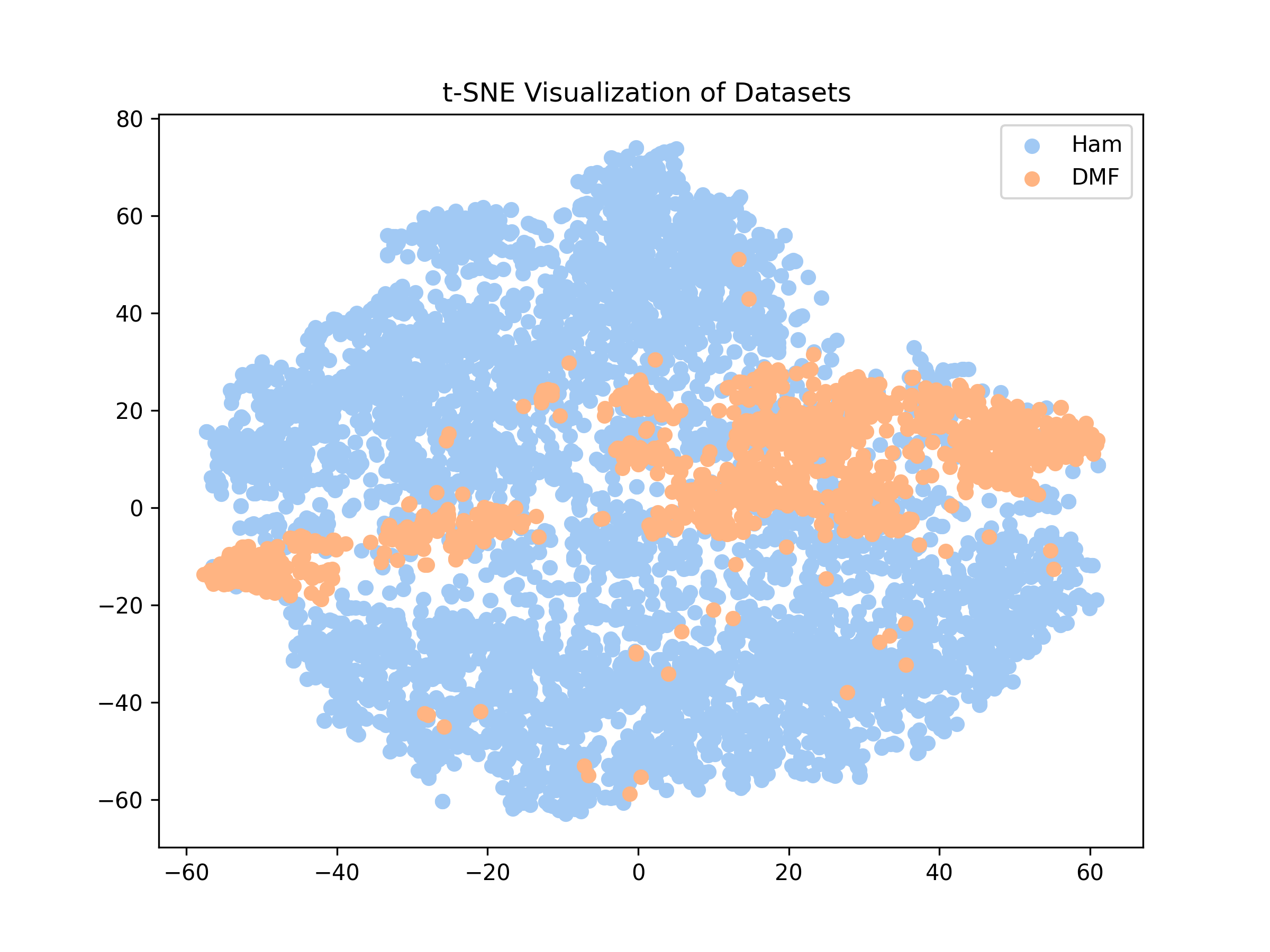} 
    \caption{A visualization of the distribution of the HAM10000 and DMF dataset through the t-SNE dimensionailty reduction method.}
    \label{fig: tsne}
\end{figure}

In the context of skin lesion classification, we have selected the DMF \cite{ballerini2013color} dataset to evaluate the performance of uncertainty quantification algorithms. The DMF dataset contains images from real patients, featuring seven different types of skin lesions. Notably, these lesion types align with those found in the HAM10000 dataset, which was utilized to train the core model. Despite having similar labels for both datasets, each dataset is captured using different medical devices in distinct facilities. This divergence in settings introduces a covariate shift in the distribution of both datasets, presenting a noteworthy challenge for the underlying model. The distribution shift is visualized through a dimensionality reduction t-SNE plot of both datasets, as depicted in Figure \ref{fig: tsne}.

\begin{figure}[h]
    \centering \includegraphics[width=1.0\linewidth]{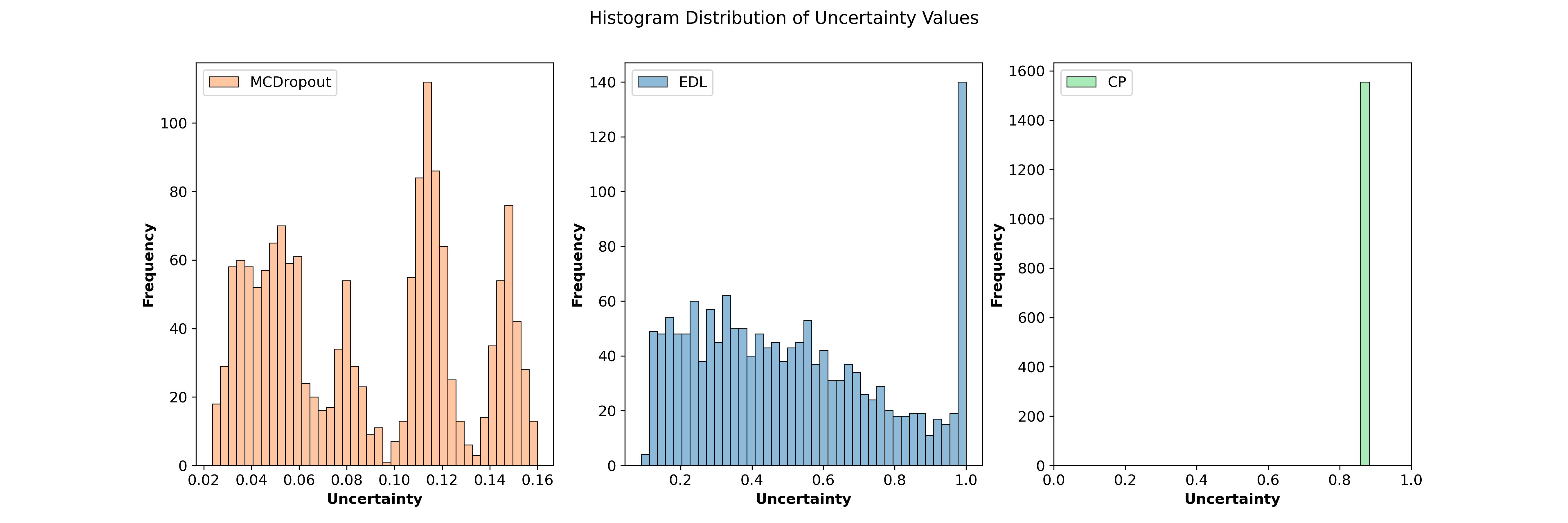} 
    \caption{Histogram distribution of uncertainty estimations by Monte Carlo dropout, evidential deep learning, and conformal prediction algorithms. A dataset with covariate shift is used to assess the performance of the three methods.}
    \label{fig: OOD}
\end{figure}

The assigned uncertainty values during inference are recorded to analyze the reaction of each method to shifted data. The average uncertainty value generated by each algorithm is reported in Table \ref{tab:average_uncertainty}, while the uncertainty distributions are shown in Figure \ref{fig: OOD}. Ideally, high uncertainty values should be assigned to samples drawn from distributions different from those used during training. Figure \ref{fig: OOD} reveals that MCD exhibits inferior performance compared to both EDL and CP. While EDL effectively assigns high uncertainty values to a larger number of samples, CP achieves the best overall performance, where all the samples are assigned a high uncertainty value, indicating that the network is highly uncertain with the predictions. 

\begin{table}[h]
    \centering
    \begin{tabular}{|c|c|}
        \hline
        \textbf{UQ algorithm} & \textbf{Average Uncertainty} \\
        \hline
        MCD & 0.86 \\
        \hline
        EDL & 0.51 \\
        \hline
        CP & 0.08 \\
        \hline
    \end{tabular}
    \caption{Average Uncertainty Values for Different uncertainty quantification algorithms. The core network is trained on HAM10000 dataset and evaluated on DMF dataset as an OOD example. Results illustrate that CP achieves the best performance in terms of OOD detection.}
    \label{tab:average_uncertainty}
\end{table}

\section{Conclusion}\label{sec: sec4}
This paper provides a comprehensive analysis of three uncertainty quantification techniques applied in the field of skin lesion medical image classification: Monte Carlo dropout, evidential deep learning, and conformal prediction. The performance of these methods is compared both qualitatively and quantitatively under various experimental conditions, to evaluate their effectiveness in quantifying uncertainty for a safety-critical application. 

CP stands out as an emerging method that provides a robust and straightforward approach to quantifying predictive uncertainty. CP generates sets of predictions, where larger sets indicate higher uncertainty and vice versa. Experimental results demonstrate CP's robustness when evaluated on correctly classified examples, showcasing its ability to generate varying levels of uncertainties that adapt to the difficulty of a given example. Notably, CP maintains consistent performance when assessed on wrongly classified examples, assigning higher uncertainty values compared to correctly classified samples. In contrast, both MCD and EDL lack these properties, leading to a superior performance of CP, as indicated in the comparative results. From an implementation point of view, CP offers lighter and more straightforward requirements compared to EDL or MCD. Unlike EDL, there is no need to retrain the core network, and unlike MCD, there is no requirement to modify the architecture. CP only requires the holdout of a dataset portion for calibration. Consequently, the method remains both model and dataset-agnostic.

In CP, the assumption that the calibration dataset is drawn from an independent and identical distribution (i.i.d) as the training dataset poses a limitation, particularly in real-world scenarios where models encounter diverse data. This assumption becomes restrictive as models are exposed to a broad spectrum of data, making the i.i.d assumption challenging to consistently uphold. While the reported results obtained from the OOD experiment showcase superior CP performance compared to other methods, it's noteworthy that the testing dataset used doesn't exhibit significant diversity from the training dataset. This limitation prompts consideration for future research directions, emphasizing the exploration of domain adaptation and generalization techniques to enhance model performance under domain shifts. In addition, it is worth acknowledging that the presence of an imbalanced dataset can present challenges that reflect real-world situations. Specifically, the effectiveness of uncertainty quantification methods is intricately tied to the performance of the core network. Therefore, it's essential to recognize that a naturally balanced dataset could potentially enhance the overall performance of these approaches.

\section*{Declaration of competing interest}
The authors declare that they have no known competing financial
interests or personal relationships that could have appeared to influence
the work reported in this paper.

\section*{Data availability}
All data and scripts are available at: \href{https://github.com/jfayyad/UQ}{https://github.com/jfayyad/UQ}

\bibliographystyle{elsarticle-num}

\bibliography{cas-refs}

\begin{thebibliography}{10}
\expandafter\ifx\csname url\endcsname\relax
  \def\url#1{\texttt{#1}}\fi
\expandafter\ifx\csname urlprefix\endcsname\relax\def\urlprefix{URL }\fi
\expandafter\ifx\csname href\endcsname\relax
  \def\href#1#2{#2} \def\path#1{#1}\fi

\bibitem{fayyad2020deep}
J.~Fayyad, M.~A. Jaradat, D.~Gruyer, H.~Najjaran, Deep learning sensor fusion
  for autonomous vehicle perception and localization: A review, Sensors 20~(15)
  (2020) 4220.

\bibitem{wiley2018computer}
V.~Wiley, T.~Lucas, Computer vision and image processing: a paper review,
  International Journal of Artificial Intelligence Research 2~(1) (2018)
  29--36.

\bibitem{alijani2022ensemble}
S.~Alijani, J.~Tanha, L.~Mohammadkhanli, An ensemble of deep learning
  algorithms for popularity prediction of flickr images, Multimedia Tools and
  Applications 81~(3) (2022) 3253--3274.

\bibitem{salari2022object}
A.~Salari, A.~Djavadifar, X.~Liu, H.~Najjaran, Object recognition datasets and
  challenges: A review, Neurocomputing 495 (2022) 129--152.

\bibitem{zhu2017deep}
X.~X. Zhu, D.~Tuia, L.~Mou, G.-S. Xia, L.~Zhang, F.~Xu, F.~Fraundorfer, Deep
  learning in remote sensing: A comprehensive review and list of resources,
  IEEE geoscience and remote sensing magazine 5~(4) (2017) 8--36.

\bibitem{lambert2022trustworthy}
B.~Lambert, F.~Forbes, A.~Tucholka, S.~Doyle, H.~Dehaene, M.~Dojat, Trustworthy
  clinical ai solutions: a unified review of uncertainty quantification in deep
  learning models for medical image analysis, arXiv preprint arXiv:2210.03736
  (2022).

\bibitem{fayyad2023out}
J.~Fayyad, Out-of-distribution detection using inter-level features of deep
  neural networks, Ph.D. thesis, University of British Columbia (2023).

\bibitem{ghesu2021quantifying}
F.~C. Ghesu, B.~Georgescu, A.~Mansoor, Y.~Yoo, E.~Gibson, R.~Vishwanath,
  A.~Balachandran, J.~M. Balter, Y.~Cao, R.~Singh, et~al., Quantifying and
  leveraging predictive uncertainty for medical image assessment, Medical Image
  Analysis 68 (2021) 101855.

\bibitem{yan2023uncertainty}
J.~Yan, S.~Cai, X.~Cai, G.~Zhu, W.~Zhou, R.~Guo, H.~Yan, Y.~Wang, Uncertainty
  quantification of microcirculatory characteristic parameters for recognition
  of cardiovascular diseases, Computer Methods and Programs in Biomedicine
  (2023) 107674.

\bibitem{jahmunah2023uncertainty}
V.~Jahmunah, E.~Ng, R.-S. Tan, S.~L. Oh, U.~R. Acharya, Uncertainty
  quantification in densenet model using myocardial infarction ecg signals,
  Computer Methods and Programs in Biomedicine 229 (2023) 107308.

\bibitem{abdar2021review}
M.~Abdar, F.~Pourpanah, S.~Hussain, D.~Rezazadegan, L.~Liu, M.~Ghavamzadeh,
  P.~Fieguth, X.~Cao, A.~Khosravi, U.~R. Acharya, et~al., A review of
  uncertainty quantification in deep learning: Techniques, applications and
  challenges, Information fusion 76 (2021) 243--297.

\bibitem{wei2022mitigating}
H.~Wei, R.~Xie, H.~Cheng, L.~Feng, B.~An, Y.~Li, Mitigating neural network
  overconfidence with logit normalization, in: International Conference on
  Machine Learning, PMLR, 2022, pp. 23631--23644.

\bibitem{guo2017calibration}
C.~Guo, G.~Pleiss, Y.~Sun, K.~Q. Weinberger, On calibration of modern neural
  networks, in: International conference on machine learning, PMLR, 2017, pp.
  1321--1330.

\bibitem{sensoy2018evidential}
M.~Sensoy, L.~Kaplan, M.~Kandemir, Evidential deep learning to quantify
  classification uncertainty, Advances in Neural Information Processing Systems
  31 (2018).

\bibitem{malinin2018predictive}
A.~Malinin, M.~Gales, Predictive uncertainty estimation via prior networks,
  Advances in neural information processing systems 31 (2018).

\bibitem{feng2024trusted}
M.~Feng, K.~Xu, N.~Wu, W.~Huang, Y.~Bai, Y.~Wang, C.~Wang, H.~Wang, Trusted
  multi-scale classification framework for whole slide image, Biomedical Signal
  Processing and Control 89 (2024) 105790.

\bibitem{kononenko1989bayesian}
I.~Kononenko, Bayesian neural networks, Biological Cybernetics 61~(5) (1989)
  361--370.

\bibitem{blei2017variational}
D.~M. Blei, A.~Kucukelbir, J.~D. McAuliffe, Variational inference: A review for
  statisticians, Journal of the American statistical Association 112~(518)
  (2017) 859--877.

\bibitem{brooks1998markov}
S.~Brooks, Markov chain monte carlo method and its application, Journal of the
  royal statistical society: series D (the Statistician) 47~(1) (1998) 69--100.

\bibitem{leibig2017leveraging}
C.~Leibig, V.~Allken, M.~S. Ayhan, P.~Berens, S.~Wahl, Leveraging uncertainty
  information from deep neural networks for disease detection, Scientific
  reports 7~(1) (2017) 1--14.

\bibitem{lakshminarayanan2017simple}
B.~Lakshminarayanan, A.~Pritzel, C.~Blundell, Simple and scalable predictive
  uncertainty estimation using deep ensembles, Advances in neural information
  processing systems 30 (2017).

\bibitem{shafer2008tutorial}
G.~Shafer, V.~Vovk, A tutorial on conformal prediction., Journal of Machine
  Learning Research 9~(3) (2008).

\bibitem{vovk2005algorithmic}
V.~Vovk, A.~Gammerman, G.~Shafer, Algorithmic learning in a random world,
  Vol.~29, Springer.

\bibitem{angelopoulos2020uncertainty}
A.~Angelopoulos, S.~Bates, J.~Malik, M.~I. Jordan, Uncertainty sets for image
  classifiers using conformal prediction, arXiv preprint arXiv:2009.14193
  (2020).

\bibitem{angelopoulos2022image}
A.~N. Angelopoulos, A.~P. Kohli, S.~Bates, M.~Jordan, J.~Malik, T.~Alshaabi,
  S.~Upadhyayula, Y.~Romano, Image-to-image regression with distribution-free
  uncertainty quantification and applications in imaging, in: International
  Conference on Machine Learning, PMLR, 2022, pp. 717--730.

\bibitem{lu2022fair}
C.~Lu, A.~Lemay, K.~Chang, K.~H{\"o}bel, J.~Kalpathy-Cramer, Fair conformal
  predictors for applications in medical imaging, in: Proceedings of the AAAI
  Conference on Artificial Intelligence, Vol.~36, 2022, pp. 12008--12016.

\bibitem{tschandl2018ham10000}
P.~Tschandl, C.~Rosendahl, H.~Kittler, The ham10000 dataset, a large collection
  of multi-source dermatoscopic images of common pigmented skin lesions,
  Scientific data 5~(1) (2018) 1--9.

\bibitem{ballerini2013color}
L.~Ballerini, R.~B. Fisher, B.~Aldridge, J.~Rees, A color and texture based
  hierarchical k-nn approach to the classification of non-melanoma skin
  lesions, Color medical image analysis (2013) 63--86.

\bibitem{he2016deep}
K.~He, X.~Zhang, S.~Ren, J.~Sun, Deep residual learning for image recognition,
  in: Proceedings of the IEEE conference on computer vision and pattern
  recognition, 2016, pp. 770--778.

\bibitem{caldeira2020deeply}
J.~Caldeira, B.~Nord, Deeply uncertain: comparing methods of uncertainty
  quantification in deep learning algorithms, Machine Learning: Science and
  Technology 2~(1) (2020) 015002.

\bibitem{gal2016dropout}
Y.~Gal, Z.~Ghahramani, Dropout as a bayesian approximation: Representing model
  uncertainty in deep learning, in: international conference on machine
  learning, PMLR, 2016, pp. 1050--1059.

\bibitem{grabinski2022robust}
J.~Grabinski, P.~Gavrikov, J.~Keuper, M.~Keuper, Robust models are less
  over-confident, Advances in Neural Information Processing Systems 35 (2022)
  39059--39075.

\bibitem{shafer1992dempster}
G.~Shafer, Dempster-shafer theory, Encyclopedia of artificial intelligence 1
  (1992) 330--331.

\bibitem{angelopoulos2021gentle}
A.~N. Angelopoulos, S.~Bates, A gentle introduction to conformal prediction and
  distribution-free uncertainty quantification, arXiv preprint arXiv:2107.07511
  (2021).

\bibitem{he2023survey}
W.~He, Z.~Jiang, A survey on uncertainty quantification methods for deep neural
  networks: An uncertainty source perspective, arXiv preprint arXiv:2302.13425
  (2023).

\bibitem{fayyad2023exploiting}
J.~Fayyad, K.~Gupta, N.~Mahdian, D.~Gruyer, H.~Najjaran, Exploiting classifier
  inter-level features for efficient out-of-distribution detection, Image and
  Vision Computing (2023) 104897.

\end{thebibliography}

\end{document}